\documentclass[12pt,leqno]{article}
\usepackage{pdflscape}
\usepackage[justification=centering]{caption}
\usepackage{ amssymb }
\newtheorem{thm}{Theorem}[section]
\newtheorem{prop}[thm]{Proposition}
\newtheorem{lem}[thm]{Lemma}
\newtheorem{cor}[thm]{Corollary}

\newcommand{\pf}{{\bf Proof. \ }}
\newcommand{\qed}{\hfill $\Box$ \\}
\font\msbm=msbm10 at 12pt

\newcommand{\FF}{\mbox{\msbm F}}

\newcommand{\vv}{{\bf v}}
\newcommand{\vw}{{\bf w}}

\begin{document}

\title{The combinatorics of LCD codes: \\Linear Programming bound and orthogonal matrices }
\author{
Steven T. Dougherty\thanks{ Department of Mathematics, University of
Scranton, Scranton, PA 18510, USA.  {Email: \tt
prof.steven.dougherty@gmail.com}}, \and Jon-Lark Kim\thanks{ Department of Mathematics,
Sogang University,
Seoul 121-742, South Korea.
{Email: \tt jlkim@sogang.ac.kr}}
\and Buket Ozkaya \thanks{ CNRS/LTCI, UMR 5141, T\'el\'ecom-ParisTech,
46 rue Barrault 75\,634 Paris cedex 13, France.
{Email: \tt ozkaya@enst.fr}}
\and Lin Sok  \thanks{
Department of Mathematics, Royal University of Phnom Penh,
Russian Federation Blvd, Phnom Penh, Cambodia.
{E-mail:\tt sok.lin@rupp.edu.kh}}
 \and Patrick Sol\'e\thanks{ CNRS/LTCI, UMR 5141, T\'el\'ecom-ParisTech,
46 rue Barrault 75\,634 Paris cedex 13, France.
{Email: \tt sole@enst.fr}}}
\date{}
\maketitle

\vspace*{3cm}

\begin{abstract}
Linear Complementary Dual codes (LCD) are binary linear codes that meet their dual trivially.
We construct LCD codes using orthogonal matrices, self-dual codes,  combinatorial designs and  Gray map from codes over the family of rings $R_k$.
We give a linear programming bound on the largest size of an LCD code of given length and minimum distance.
We make a table of lower bounds for this combinatorial function for modest values of the parameters.
\end{abstract}

\vspace*{1cm}

\bf Key Words\rm : LCD (linear codes with complementary dual) codes, Self-dual codes, Linear programming bound.

\bf MSC (2010) 94B 05, 20H 30

\vspace{1.5cm}

\newpage

%%%%%%%%%%%%%%%%%%%%%%%%%%%%%%%%%%%%%%%%%
\section{Introduction}

In this paper, we study linear codes with complementary duals, which we refer to as LCD codes. 
These codes were introduced by Massey in \cite{Massey} and  give an optimum linear coding solution for the two user binary adder channel.
They are also used in counter measures to passive and active side channel analyses on embedded cryto-systems, see \cite{CarGui} for a detailed description.   

The main result is a linear programming bound on the largest size of an LCD code of given length and minimum distance. We show by numerical examples
that this bound is, in general, sharper than the standard linear programming bound on the size of codes of given length and distance. We also give a combinatorial construction of LCD codes based on orthogonal matrices,
which are essentially equivalent to systematic generator matrices of self-dual codes. They also enjoy a pseudo-random construction due to their multiplicative groups
structure. It is important to note that a single self-dual code, or, equivalently a single orthogonal matrix  give rise to several LCD codes.
We sketch another construction by codes over rings and Gray maps, and a construction based on symmetric designs. A table of lower bounds on the largest 
size of an LCD code of given length and minimum distance is built based on the orthogonal matrix construction.

The material is organized as follows. Section 2 contains some constructions of LCD codes over rings that impact LCD codes over fields. Section 3 introduces
and studies
two combinatorial functions related to LCD codes. Section 4 derives the linear programming bound 
and provide a comparative numerical table with the standard linear programming bound. Section 5 contains the various constructions from rings, matrices
and block designs. A last section concludes the paper and paves the way to new research.
%%%%%%%%%%%%%%%%%%%%%%%%%%%%%%%%%%%%%%%%%%%%%%%%%%%%%%%%%
\section{Preliminaries}

In this work, we shall be largely concerned with codes over finite fields.  However, we shall use codes over rings together with a linear Gray map to construct LCD codes.  Hence we shall make the definitions in a general setting.  A code $C$ of length $n$ over a ring $R$ is a subset of $R^n.$  All rings in this paper are assumed to be commutative rings with unity.  If the code is  a submodule then the code is said to be linear.  Attached to the ambient space is the standard inner-product, namely $[\vv,\vw ] = \sum v_i w_i.$  The orthogonal is defined by $C^\perp = \{ \vv \in R^n \ | \ [\vv,\vw]=0,\ \forall \vw \in C \}.$ If $R$ is a Frobenius ring then we have that $|C||C^\perp| = |R|^n$. For codes over finite fields we have $dim(C) + dim(C^\perp) = n.$  

A linear code with complementary code (LCD) is a linear code $C$ satisfying $C \cap C^\perp = \{ {\bf 0} \}.$
Any code over a field is equivalent to a code generated by a matrix of the form $( I_k \ | A)$ where $I_k$ denotes the $k$ by $k$ identity matrix.  For codes over rings, this is not the case so we shall talk about generating vectors instead in the following lemmas.

    \begin{lem}\label{firstlem}
    Let $\vv_1,\vv_2,\dots,\vv_k$ be a vectors over a commutative ring of characteristic 2  such that $[\vv_i, \vv_i]=1$  for each $i$ and $[\vv_i,\vv_j]=0$ for $i \neq j.$ Then 
    $C = \langle  \vv_1,\vv_2,\dots,\vv_k \rangle $ is an LCD code.
    \end{lem}
    \pf
    Any vector in $C$ is of the form $\vw= \sum_{i \in A} \vv_i.$  Then let $j \in A$, it follows that $[\vv_j,\vw ] =1.  $ Hence $\vw \not \in 
    C^\perp.$ This gives that no non-trivial element in $C$ is also in $C^\perp$ and hence their intersection is trivial.
\qed

Applying this lemma to codes over fields we have the following.   
    
  \begin{cor}\label{cor1}
Let $G$ be a generator matrix for a code over a finite field. 
If $GG^T = I_n$ then $G$ generates an LCD code. 
\end{cor}  
    
    More generally for codes over fields this leads to the following.
    
    \begin{cor}\label{cor2}
    Let $G$ be a generator matrix for a code over a field.  Then $det(GG^T) \neq 0$ if and only if $G$ generates an LCD code.
    \end{cor}

 \begin{lem} \label{secondlem}  
Let $\vv_1,\vv_2,\dots,\vv_k$ be a set of vectors over a ring of characteristic 2 such that $[\vv_i,\vv_i] = 0$ and $[\vv_i,\vv_j ] =1$ if $i \neq j.$
Then $C = \langle \vv_1,\vv_2,\dots,\vv_k \rangle $ is LCD if and only if $k$ is even.
\end{lem}
\pf
Assume $k$ is even.
Consider the vector $\vw = \sum_{i \in A} \vv_i.$

If $|A|$ is even take $j \in A$.  Then $[\vw,\vv_j] = 1.$  If $|A|$ is odd take $j \not \in A$  then $[\vw,\vv_j]=1.$ In either case, no linear combination of the generators can be in the orthogonal.  Hence the code is LCD.

Assume $k$ is odd.  Then $[\sum_{i=1}^k  \vv_i, \vv_j] = 0$ for any $j$.  Hence $\sum_{i=1}^k  \vv_i \in C \cap C^\perp$ and the code is not LCD.
\qed

Let $J_n$ denote the all one $n$ by $n$ matrix.
Considering this lemma as applied to codes over fields we have the following.

\begin{cor} \label{cor3} 
Let $G$ be a generator matrix for a code over a finite field. 
If $GG^T = J_n -I_n$, $n$ even,  then $G$ generates an LCD code. 
\end{cor}

%%%%%%%%%%%%%%%%%%%%%%%%%%%%%%%%%%%%%%%%%%%%%%%%%%%%%%%%
\section{Elementary Bounds}
In this section, we are only concerned with codes over the binary field.
\subsection{Fixed $n$ and $k.$}

Let LCD$[n,k] := \max\{ d ~|~ {\mbox{there exists a binary }} [n,k,d] {\mbox{ LCD code}} \}.$

\begin{lem} \label{lem_length_n}
For $n$ and $k$ integers greater than 0, LCD$[n+1,k] \ge$ LCD$[n,k]$.
\end{lem}

\pf
Let $G$ be a generator matrix of an $[n,k,d]$  LCD code $C$. Then $GG^T$ is invertible since $C$ is LCD. Let $\bar{G}$ be the matrix obtained from $G$ by adding the zero column {\bf 0} to the right end of $G$, that is, $\bar{G} = G{\bf 0}$. $\bar{G} (\bar{G})^T = GG^T$ is invertible. Hence $\bar{G}$ generates an $[n+1,k,d]$ LCD code. Therefore LCD$[n+1,k] \ge$ LCD$[n,k]$.
\qed

\begin{prop}

\begin{itemize}
\item[{(i)}] If $n$ is odd, then LCD$[n,1]=n$ and LCD$[n, n-1]=2$.

\item[{(ii)}] If $n$ is even, then LCD$[n,1]=n-1$ and LCD$[n, n-1]=1$.
\end{itemize}

\end{prop}
 
 %\begin{pf}
 \pf
 (i) It is clear that the repetition $[n,1,n]$ code and its dual are LCD and have the highest minimum distances.

(ii) If $n$ is even, the repetition $[n,1,n]$ code is not LCD since its dual contains the all-one vector. It is easy to see that the code $C$ with generator matrix $[0~1~1\dots 1]$ is LCD. Thus LCD$[n,1]=n-1$. The dual $C^{\perp}$ of $C$ is LCD with minimum distance 1. If LCD$[n, n-1]=2$, then the corresponding code is the even $[n, n-1,2]$ code which is not LCD since the all-one vector belongs to the even code and the repetition code of length $n$. Thus LCD$[n, n-1]=1$.
\qed

\begin{lem}\label{lem_prod_sum} The following hold.

\begin{enumerate}

\item[{(i)}]
LCD$[nm,kl] \ge$ LCD$[n,k]$ LCD$[m,l]$.

\item[{(ii)}]
LCD$[n+m,k+l] \ge$ $\min$\{{LCD$[n,k]$, LCD$[m,l]$\}}.
\end{enumerate}

\end{lem}

%\begin{pf}
\pf
(i) Let $G_1$ be a generator matrix of an $[n,k,d_1]$ LCD code $C_1$ and $G_2$ a generator matrix of an $[m, l, d_2]$ LCD code $C_2$. Consider the direct product of $C_1$ and $C_2$, denoted by $C_1 \otimes C_2$, which has parameters $[nm, kl, d_1 d_2]$ (see~\cite[Ch. 8]{MS}). The generator matrix of $C_1 \otimes C_2$ is the Kronecker product of $G_1$ and $G_2$, denoted by $G_1 \otimes G_2$. We need to show that $C_1 \otimes C_2$ is LCD. It suffices to show that $(G_1 \otimes G_2) (G_1 \otimes G_2)^T$ is invertible. Note that $(G_1 \otimes G_2) (G_1 \otimes G_2)^T = (G_1 \otimes G_2) (G_1^T \otimes G_2^T)
=(G_1 G_1^T \otimes G_2 G_2^T)$. Since $G_iG_i^T$ (for $i=1,2$) is invertible, $(G_1 G_1^T \otimes G_2 G_2^T)$ is invertible because $(G_1 G_1^T \otimes G_2 G_2^T) ((G_1 G_1^T)^{-1} \otimes (G_2 G_2^T)^{-1})=
I_k \otimes I_l=I_{kl}$, where $I_a$ is the identity matrix of order $a$.

(ii) It is known~\cite{CarGui} that the direct sum of $C_1 \oplus C_2$ of LCD codes $C_1, C_2$ of parameters $[n, k, d_1]$ and $[m, l, d_2]$ respectively is also an LCD code of parameters $[n+m, k+l, \min\{d_1, d_2 \}]$. Hence LCD$[n+m,k+l] \ge$ $\min$\{{LCD$[n,k]$, LCD$[m,l]$\}} follows.
\qed
%\end{pf}

%%%%%%%%%%%%%%%%%%%%%%%%%%%%%%%%%%%%%%%%%%%
\subsection{ LCD $[n,k]$ for small $n,k$}\label{small}

We have a partial result on LCD$[n,2]$ for $n \ge 3$.

\begin{thm} We have the following:

\begin{enumerate}

\item[{(i)}] LCD$[3,2]=2$

\item[{(ii)}] LCD$[4,2]=2$

\item[{(iii)}] LCD$[5,2]=2$

\item[{(iv)}] LCD$[6,2]=3$

\item[{(v)}] LCD$[7,4]=2$

\end{enumerate}
\end{thm}
\begin{pf}
\begin{enumerate}
 \item Choose the even code of length $3$ with generator matrix $G=\left[ \begin{array}{ccc}
                  1 & 1 &0 \\
                  1 & 0 &1\\
               \end{array}
               \right]$.
               It is LCD and has minimum distance $d=2$. There is no $[3,2,3]$ code. Thus LCD$[3,2]=2$.
               
\item  There is $[4,2,2]$ LCD namely the $BKLC(GF(2),4,2)$ provided by Magma. Hence LCD$[4,2]=2$.

 \item  Since there is no nontrivial MDS binary code, there is no $[5,2,4]$ code. There are two $[5,2,3]$ codes up to equivalence. They have generator matrices such as
  $$\left[ \begin{array}{ccccc}
                  1 & 0 & 1 & 1& 0 \\
                  0 & 1 & 1 & 1& 1\\
               \end{array}
               \right] {\mbox{~and}}
 \left[ \begin{array}{ccccc}
                  1 & 0 & 1 & 1 & 0 \\
                  0 & 1 & 0 & 1 & 1 \\
               \end{array}
               \right].
  $$
 None of them are LCD. Thus LCD$[5,2] \le 2$, and so LCD$[5,2]=2$ by Lemma~\ref{lem_length_n} and (i) above.

 \item  Note that the repetition $[3,1,3]$ code is LCD. Thus by (ii) of Lemma~\ref{lem_prod_sum}, we have LCD$[6,2] \ge 3$. If LCD$[6,2]=4$, then there is a unique $[6,2,4]$ code $C$ whose generator matrix can be arranged up to equivalence as 
$$G=\left[ \begin{array}{cccccc}
                  1 & 0 & 1 & 1& 1 & 0 \\
                  0 & 1 & 0 & 1 & 1& 1\\
               \end{array}
               \right].$$
Since $GG^T$ is a zero matrix (hence noninvertible), $C$ is not LCD. Thus LCD$[6,2]=3$.

\item There is a $[7,2,4]$  code whose generator matrix is given by
 $$G=\left[ \begin{array}{ccccccc}
                  1 & 1 & 1 & 1& 0 & 0 & 0\\
                  0 & 0 & 0 & 1& 1 & 1 & 1\\
               \end{array}
               \right].$$
It is known that an optimal $[7,2]$ code has $d=4$~\cite{Gra}. Therefore LCD$[7,2]=4$.
\end{enumerate}
 \end{pf}

\subsection{Fixed $n$ and $d$}
We introduce the combinatorial function LCK$[n,d] := \max\{ k ~|~ {\mbox{there exists a binary }} [n,k,d] {\mbox{ LCD code}} \}.$
From first principles, we see that LCK$[n,d]\le \log_2 A(n,d).$
Some values of that function for small $d$ are easy to find.
\begin{prop}
 For all $n\ge 1,$ we have LCK$[n,1]= n.$ 
\end{prop}

\pf
 The complete code with dual the null code is LCD. The result follows.
\qed

\begin{prop}
 If $n$ is odd then LCK$[n,n]=1$ and LCK$[n,2]=n-1.$
 If $n$ is even then LCK$[n,n]=0,$ and LCK$[n,2]= n-2.$
\end{prop}

\pf
 The repetition code of odd length is $LCD$ and optimal.  Its dual is as well. The first assertion follows.
 On the other hand the repetition code of even length is unique with their parameters and self-orthogonal, hence not LCD.
 By adding an extra zero/one coordinate to the dual of the repetition code in length $n-1$ we obtain an LCD code of parameters $[n,n-2,2].$
\qed

\begin{prop}
 For all integers $m>1$ we have 
 \begin{itemize}
  \item LCK$[2^m-1,3]<2^m-m-1$
  \item LCK$[2^m-1,2^{m-1}]<m.$
 \end{itemize}

\end{prop}
\pf
 The Hamming code and its dual the Simplex code are not LCD since the Simplex code is self-orthogonal. Further these codes are unique with their parameters.
 This is immediate for the Hamming code. Note that the Simplex code is unique as meeting the Plotkin bound \cite[Th. 11(a)]{MS}, hence one-weight, 
 hence characterized by \cite{B}.
\qed

\begin{prop}
We have
LCK$[24,8]<12,$
and 
LCK$[23,7]<11.$
\end{prop}
\pf
 The extended Golay code is unique \cite[Th. 104]{P} and self-dual.
 Puncturing once yields another unique code \cite[Th. 104]{P} which contains its dual.
\qed

Let $g(k,d)=\sum_{i=0}^{k-1} \lceil \frac{d}{2^i} \rceil$ denote the RHS of the Griesmer bound.
\begin{prop}
If $ d$ is a multiple of four, then LCK$[g(k,d),d]<k.$
\end{prop}
\pf
 By Theorem 1 of \cite{W}, if a code $C$ meets the Griesmer bound with a minimum distance multiple of $4,$ then all its weight are multiples of $4.$ It is then 
 immediate by the parallelogram identity \cite[(12) p.9]{MS}
 that $C$ is self-orthogonal.
\qed
\section{Linear Programming bound }
%\subsection{Derivation of the bound}
Let $C$ denote a binary linear $[n,k]$ code and $A_i$ its weight distribution.
Let $B_i$ denote the weight distribution of its dual code $C^{\perp}.$
Let $P_i(x)$ be the Krawtchouk polynomial of degree $i$ given by the following generating function:

$$ \sum_{i=0}^nP_i(x)z^i= (1+z)^{n-x}(1-z)^x.$$
If $M$ is a square matrix of order $N_R$ by $N_C$ and $x,h$ are column vectors of length $N_R,$ and $N_C$ respectively, we denote by
 $U(M,h)$ the maximum of $\sum_{j=1 }^{N_C} x_i$ for  nonnegative rationals $x_i$ subjected to the $N_R$ linear constraints
 $Mx \le h.$
 We need the following auxiliary matrices:
 \begin{itemize}
  \item $P=(P_j(i))$;
  \item $\Delta$ the matrix with entries $\Delta_{i,j}={n \choose i }$ for all $1\le j \le n;$
  \item $I_m$ the identity matrix of order $m.$
 \end{itemize}

If $C$ is LCD then for all $n\ge i\ge 1$ we have from the definition of $LCD$ codes that
 \begin{equation}\label{defi}
  A_i+B_i\le {n \choose i },
 \end{equation}

a vector of weight being in either $C$ or its dual but not in both.

Now by the MacWilliams formula we know that
\begin{equation}\label{mac}
B_i=2^{-k} \sum_{j=0}^n A_jP_i(j).
\end{equation}
 Writing $2^k=\sum_{i=0}^n A_i,$ we get the following bound.
 
 {\prop If $k\ge k_0,$ then for all $n\ge i\ge 1$ we have
 $$ 2^{k_0}A_i\le \sum_{j=1}^n A_j({n \choose i }- P_i(j)).$$ 
 }
 \pf
 We eliminate $B_i$ between equation (\ref{defi}) premultiplied by $2^k$ and (\ref{mac}) and rearrange. To avoid quadratic terms we bound $2^k A_i$ below by $ 2^{k_0}A_i.$
 Note that by the generating function for Krawtchouk polynomials $P_i(0)={n \choose i }.$
 \qed

 We consider the block matrix $M(n,k_0,d)$ of order $2n+d-1$ by $n$ with successive block rows 
 $I_{d-1},\, P-\Delta+2^{k_0}I_n,\,-P.$
 {\thm  If $k\ge k_0,$  then $$2^{LCK[n,d]} \le 1+U(M(n,k_0,d),0).$$ }
 
 \pf
 The three type of constraints come from, in order, the distance of $C,$ the above proposition and the Delsarte inequalities
 (nonnegativity of the $B_i$'s).
 \qed

 %\includepdf[pages={1}]{joint table.pdf}
\clearpage
\begin{landscape}
\begin{table}[ht]
\centering \resizebox{1.4\textheight}{!}{
\begin{tabular}{|c||c|c|c|c|c|c|c|c|c|c|c|c|c|c|c|c|c|c|c|c|c|c|c|c|c|c|c|c|c|c|}
  \hline
  \textbf{n/d} & \textbf{1} & \textbf{2} & \textbf{3} & \textbf{4} & \textbf{5} & \textbf{6} & \textbf{7} & \textbf{8} & \textbf{9} & \textbf{10} & \textbf{11} & \textbf{12} & \textbf{13} & \textbf{14} & \textbf{15} & \textbf{16} & \textbf{17} & \textbf{18} & \textbf{19} & \textbf{20} & \textbf{21} & \textbf{22} & \textbf{23} & \textbf{24} & \textbf{25} & \textbf{26} & \textbf{27} & \textbf{28} & \textbf{29} & \textbf{30}
  \\\hline \hline
  \textbf{1} & 1 &  &  &  &  &  &  &  &  &  &  &  &  &  &  &  &  &  &  &  &  &  &  &  &  &  &  &  &  &  \\\hline
  \textbf{2} & 2 & 0(1) &  &  &  &  &  &  &  &  &  &  &  &  &  &  &  &  &  &  &  &  &  &  &  &  &  &  &  &  \\\hline
  \textbf{3} & 3 & 2 & 1 &  &  &  &  &  &  &  &  &  &  &  &  &  &  &  &  &  &  &  &  &  &  &  &  &  &  &  \\\hline
  \textbf{4} & 4 & 2(3) & \underline{1} & 0(1) &  &  &  &  &  &  &  &  &  &  &  &  &  &  &  &  &  &  &  &  &  &  &  &  &  &  \\\hline
  \textbf{5} & 5 & 4 & 2 & \underline{1} & 1 &  &  &  &  &  &  &  &  &  &  &  &  &  &  &  &  &  &  &  &  &  &  &  &  &  \\\hline
  \textbf{6} & 6 & 4(5) & 3 & 2 & \underline{1} & 0(1) &  &  &  &  &  &  &  &  &  &  &  &  &  &  &  &  &  &  &  &  &  &  &  &  \\\hline
  \textbf{7} & 7 & 6 & 4 & 3 & \underline{1} & \underline{1} & 1 &  &  &  &  &  &  &  &  &  &  &  &  &  &  &  &  &  &  &  &  &  &  &  \\\hline
  \textbf{8} & 8 & 6(7) & \underline{4} & 3(4) & 2 & \underline{1} & \underline{1} & 0(1) &  &  &  &  &  &  &  &  &  &  &  &  &  &  &  &  &  &  &  &  &  &  \\\hline
  \textbf{9} & 9 & 8 & 5 & 4 & \underline{2} & 2 & \underline{1} & \underline{1} & 1 &  &  &  &  &  &  &  &  &  &  &  &  &  &  &  &  &  &  &  &  &  \\\hline
  \textbf{10} & 10 & 8(9) & 6 & 5 & 3 & 2 & \underline{1} & \underline{1} & \underline{1} & 0(1) &  &  &  &  &  &  &  &  &  &  &  &  &  &  &  &  &  &  &  &  \\\hline
  \textbf{11} & 11 & 10 & 7 & 6 & 4 & 3 & 2 & \underline{1} & \underline{1} & \underline{1} & 1 &  &  &  &  &  &  &  &  &  &  &  &  &  &  &  &  &  &  &  \\\hline
  \textbf{12} & 12 & 10(11) & 8 & 7 & \underline{5} & 4 & \underline{2} & 2 & \underline{1} & \underline{1} & \underline{1} & 0(1) &  &  &  &  &  &  &  &  &  &  &  &  &  &  &  &  &  &  \\\hline
  \textbf{13} & 13 & 12 & 9 & 8 & 6* & 5* & 3 & 2 & \underline{1} & \underline{1} & \underline{1} & \underline{1} & 1 &  &  &  &  &  &  &  &  &  &  &  &  &  &  &  &  &  \\\hline
  \textbf{14} & 14 & 12(13) & 10 & 9 & 7* & 6* & 4 & 3 & 2 & \underline{1} & \underline{1} & \underline{1} & \underline{1} & 0(1) &  &  &  &  &  &  &  &  &  &  &  &  &  &  &  &  \\\hline
  \textbf{15} & 15 & 14 & 11 & 10 & 8* & 7* & 5 & 4 & \underline{2} & 2 & \underline{1} & \underline{1} & \underline{1} & \underline{1} & 1 &  &  &  &  &  &  &  &  &  &  &  &  &  &  &  \\\hline
  \textbf{16} & 16 & 14(15) & \underline{11} & 10(11) & 8 & 7(8*) & \underline{5} & 4(5) & \underline{2} & 2 & \underline{1} & \underline{1} & \underline{1} & \underline{1} & \underline{1} & 0(1) &  &  &  &  &  &  &  &  &  &  &  &  &  &  \\\hline
  \textbf{17} & 17 & 16 & 12 & 11 & 9 & 8 & 6 & 5 & 3 & \underline{2} & 2 & \underline{1} & \underline{1} & \underline{1} &\underline{1} & \underline{1} & 1 &  &  &  &  &  &  &  &  &  &  &  &  &  \\\hline
  \textbf{18} & 18 & 16(17) & 13 & 12 & \underline{10} & 9 & 7 & 6 & \underline{4} & 3 & \underline{2} & 2 & \underline{1} & \underline{1} & \underline{1} & \underline{1} & \underline{1} & 0 1) &  &  &  &  &  &  &  &  &  &  &  &  \\\hline
  \textbf{19} & 19 & 18 & 14 & 13 & 11* & 10* & 8 & 7 & 5* & 4* & \underline{2} & 2 & \underline{1} & \underline{1} & \underline{1} & \underline{1} & \underline{1} & \underline{1} & 1 &  &  &  &  &  &  &  &  &  &  &  \\\hline
  \textbf{20} & 20 & 18(19) & 15 & 14 & 12* & 11* & 9 & 8 & 6* & 5* & 3 & \underline{2} & 2 & \underline{1} & \underline{1} & \underline{1} & \underline{1} & \underline{1} & \underline{1} & 0(1) &  &  &  &  &  &  &  &  &  &  \\\hline
  \textbf{21} & 21 & 20 & 16 & 15 & 12 & 12* & 10 & 9 & \underline{6} & 6* & \underline{3} & 3 & \underline{2} & 2 & \underline{1} & \underline{1} & \underline{1} & \underline{1} & \underline{1} & \underline{1} & 1 &  &  &  &  &  &  &  &  &  \\\hline
  \textbf{22} & 22 & 20(21) & 17 & 16 & 13 & 12 & 11 & 10 & 7* & 6 & 4 & 3 & \underline{2} & 2 & \underline{1} & \underline{1} & \underline{1} & \underline{1} & \underline{1} & \underline{1} & \underline{1} & 0(1) &  &  &  &  &  &  &  &  \\\hline
  \textbf{23} & 23 & 22 & 18 & 17 & 14 & 13 & 12 & 11 & 8* & 7* & 5 & 4 & \underline{2} & \underline{2} & 2 & \underline{1} & \underline{1} & \underline{1} & \underline{1} & \underline{1} & \underline{1} & \underline{1} & 1 &  &  &  &  &  &  &  \\\hline
  \textbf{24} & 24 & 22(23) & 19 & 18 & \underline{15} & 14 & \underline{12} & 11(12) & \underline{9} & 8* & \underline{6} & 5 & 3 & \underline{2} & \underline{2} & 2 & \underline{1} & \underline{1} & \underline{1} & \underline{1} & \underline{1} & \underline{1} & \underline{1} & 0(1) &  &  &  &  &  &  \\\hline
  \textbf{25} & 25 & 24 & 20 & 19 & 16* & 15* & \underline{13} & 12 & 10* & 9* & 6 & 6* & \underline{3} & 3 & \underline{2} & 2 & \underline{1} & \underline{1} & \underline{1} & \underline{1} & \underline{1} & \underline{1} & \underline{1} & \underline{1} & 1 &  &  &  &  &  \\\hline
  \textbf{26} & 26 & 24(25) & 21 & 20 & 17* & 16 & 14* & 13* & 10* & 10* & 7 & 6 & 4 & 3 & \underline{2} & \underline{2} & 2 & \underline{1} & \underline{1} & \underline{1} & \underline{1} & \underline{1} & \underline{1} & \underline{1} & \underline{1} & 0 (1) &  &  &  &  \\\hline
  \textbf{27} & 27 & 26 & 22 & 21 & 18* & 17* & 14 & 14* & 11* & 10* & \underline{8} & 7 & 5 & 4 & 3 & \underline{2} & \underline{2} & 2 & \underline{1} & \underline{1} & \underline{1} & \underline{1} & \underline{1} & \underline{1} & \underline{1} & \underline{1} & 1 &  &  &  \\\hline
  \textbf{28} & 27(28) & 26(27) & 22 & 21 & 18* & 17* & 14 & 14* & 11* & 10* & \underline{8} & 7 & 5 & 4 & 3 & \underline{2} & \underline{2} & 2 & \underline{1} & \underline{1} & \underline{1} & \underline{1} & \underline{1} & \underline{1} & \underline{1} & \underline{1} & \underline{1} & 0(1) &  &  \\\hline
  \textbf{29} & 28(29) & 27 & 24 & 23 & 20* & 19* & 16* & 15* & 13* & 12* & 10* & 9* & 7* & 6* & 4 & 3 & \underline{2} & \underline{2} & 2 & \underline{1} & \underline{1} & \underline{1} & \underline{1} & \underline{1} & \underline{1} & \underline{1} & \underline{1} & \underline{1} & 1 &  \\\hline
  \textbf{30} & 29(30) & 28(29) & 25 & 24 & 20 & 20* & 17* & 16* & 14* & 13* & 10 & 10* & \underline{7} & 7* & 5 & 4 & \underline{2} & \underline{2} & \underline{2} & 2 & \underline{1} & \underline{1} & \underline{1} & \underline{1} & \underline{1} & \underline{1} & \underline{1} & \underline{1} & \underline{1} & 0(1) \\
  \hline
\end{tabular}}
\caption{
            $(\,)$ : classical LP bound value, if different\\
            $\underline{\,}$: no such code with those parameters\\
            $^*$ :  larger than the dimension of the best known code with the given parameters
                    }
\end{table}
\end{landscape}

\section{Constructions}
%%%%%%%%%%%%%%%%%%%%%%%%%%%%%%%%%%%

\subsection{Rings}  

In this section, we shall examine a family of rings over which we can define a Gray map which can be used to construct LCD codes.

The ring $R_k$ is defined as $R_k = \FF_2[u_1,u_2,\dots,u_k] / \langle  u_i^2, u_iu_j - u_j u_i \rangle .$
The ring $R_k$ has $|R_k| = 2^{2^k}$ and
it  is a non-chain ring which has characteristic $2$ with
    maximal ideal ${\bf M}=\langle   u_1, u_2, \ldots, u_k \rangle$
    and $Soc(R_k)=\langle u_1u_2 \cdots u_k \rangle.$ 

%For a given positive integer $k$, let $u_1, u_2, \dots, u_k$ be a set of $k$ indeterminates. For any subset $A\subseteq %\{1,2,\ldots,k\}$ we denote
%by $u_A:=\prod_{i\in A}{u_i}$ with the convention that
%$u_{\emptyset}=1.$   We note that no $u_i$ can be represented more than once in this product. 

We can now construct a linear Gray map from $R_k$ to $\FF_2^{2^k}$.  Let $\phi_1$ be the
map defined on $R_1$, namely 
$\phi_1(a+bu) = (b, a+b)$.  Then let ${{c}}\in
R.$ We can write ${{c}}={{c}}_1+u_k {{c}}_2$ where
${{c}}_1,{{c}}_2$ are elements of the ring $R_{k-1}$ of order
$2^{2^{k-1}}$,  then we define
\begin{equation} \phi_k({{c}})=(\phi_{k-1}({{c}}_2),\phi_{k-1}({{c}}_1)+\phi_{k-1}({{c}}_2)).\end{equation}
 The map $\phi_k$ is a weight preserving map which we then  expand  coordinatewise to $R^n$.  

The following can be found in \cite{Rk}.

\begin{lem}
 The map $\phi_k:R_k \rightarrow \FF_2^{2^k}$ is a linear bijection. Moreover,  we have $\phi(C^\perp) = \phi(C)^\perp.$  
\end{lem}

We can define an LCD code over $R_k$ in the usual way by saying that the code is LCD if its intersection with its dual is $\{ {\bf 0 } \}.$   
This leads immediately to the following.

\begin{thm} \label{oioioi}
Let $C$ be a LCD code of length $n$ over $R_k$ then $\phi(C)$ is a binary LCD code of length $2^kn.$  
\end{thm} 
\pf
We have that $\phi(C) \cap \phi(C)^\perp = \phi(C) \cap \phi(C^\perp)$. Then since $\phi$ is a bijection and $C \cap C^\perp = {\bf 0}$ we have the desired result.
\qed

\begin{thm}
There are no non-trivial  LCD codes of length 1 over  $R_k$.
\end{thm}
\pf
Any code of length 1 is an ideal in the ring $R$ and hence $C$ and $C^\perp$ are ideals and hence both contained in the maximal ideal ${\bf M}.$  This implies their intersection contains ${\bf M}^\perp = Soc(R) = \{{\bf 0}, u_1u_2 \cdots u_k \}$ which is non-trivial. 
\qed

\begin{thm}
\begin{enumerate}
\item[{(i)}] Let $G$ be a binary matrix such that $GG^T = I_n$, then $G$ generates an LCD code $C$ of length $n$ over $R_k$ and $\phi_k(C)$ is a binary LCD code of length $2^kn.$
\item[{(ii)}] Let $G$ be a binary matrix such that $GG^T = J_n - I_n$, $n$ even, then $G$ generates an LCD code $C$ of length $n$ over $R_k$ and $\phi_k(C)$ is a binary LCD code of length $2^kn.$
\end{enumerate}
\end{thm}
\pf
The first item follows from Corollary~\ref{cor1} followed by Theorem~\ref{oioioi}.  The second item follows from Lemma~\ref{secondlem} followed by Theorem~\ref{oioioi}. 
\qed

\subsection{Orthogonal matrices}\label{orto}
One way to construct generator matrices $G$ such that $GG^T$ is invertible is to demand $GG^T=I.$   Such rectangular matrices $G$ can be obtained as row submatrices
from so-called orthogonal matrices over $\FF_2.$
Define the orthogonal group $O(n,2)$ as the set of all matrices $X$ of $GL(n,2)$ satisfying $XX^T=I_n.$
The order of this group is known to be
$$| O(n,2)|= 2^{k^2}\prod_{i=1}^k(2^{2i}-1),$$ where $k=\lfloor n/2\rfloor.$
See \cite{M}. Generators for this group are as follows. Let $P_n$ denote the matrix group of permutation matrices of order $n.$
A transvection attached to vector $u$ is a transform $T_u$ that maps all $x\in \FF_2^n$ to $T_u(x)=x+(x,u)u.$ By \cite[Th. 19]{J} we know that
for $n\ge 4$ we have $O(n,2)=\langle P_n, T_u\rangle,$ for any  $u$ of Hamming weight $4.$ Since, as is well-known, is generated by a transposition and an
$n-$cycle the group $O(n,2)$ is generated by three generators for $n\ge 4.$ It is therefore easy to generate random orthogonal matrices of order $n$ for
large $n.$ This technique was used in \cite{AGKSS,FSSW} in the contexts of self-dual codes and self-dual Boolean functions, respectively.

Another technique is to use the correspondence with systematic generator matrices of self-dual codes. Thus $(I,X)$ is self-dual if and only if  $X\in GL(n,2).$

{\bf Example:} The Golay code of length $24$ gives an orthogonal matrix of order $12,$ which, in turn, by taking the span of some rows gives LCD codes
with parameters $[12,6,3],\,[12,4,4],\,[12,8,2].$
%%%%%%%%%%%%%%%%%%%%%%%%%%%%%%%%%%%%%%%%%%%%%%%%%

\subsection{Block designs}

Recall that for a Balanced Incomplete Block Design (BIBD) with parameters $(b,v,k,r,\lambda)$, the $b$ indicates the size of the blocks, $v$ indicates the number of varieties, $k$ indicates the number of varieties on a block, $r$ indicates the number of blocks through a variety and through any 2 varieties there are $\lambda $ blocks. We refer to this BIBD as a $2-(v,k,\lambda)$  design.

\begin{thm}
Denote by $Q$ the variety vs block incidence matrix of a $2-(v,k,\lambda)$ BIBD.
If $rk(r-\lambda)\neq 0 \pmod{2}$ then $Q$ generates an LCD code.  
\end{thm}
\begin{pf}
 It is well-known, \cite[Th. 1.4.1]{AK}, that 
$$\det(QQ^T)=rk(r-\lambda)^{v-1}.$$ Thus, provided that $rk(r-\lambda)\neq 0 \pmod{2},$ we see that the row span of $Q$ is LCD of parameters
$[b,v,\ge 2(r-\lambda)]$ by Corollary~\ref{cor2}.
\end{pf}

%%%%%%%%%%%%%%%%%%%%%%%%%%%%%%%%%%%%%%%%%%%%%
%%%%%%%%%%%%%%%%%%%%%%%%%%%%%%%%%%%%%%%
\subsection{Table of lower bounds on LCK$[n,d]$}
The first seven rows of the following table were filled up using the codes in Section \ref{small} and the results in the following section.
The remaining rows of the following table were filled up using orthogonal matrices constructed from
\begin{itemize}
 \item extended quadratic residue codes for $n=12,16,24;$
 \item database of self-dual codes \cite{jap} for the other $n=8,9,10,11,13,14,15,18,20;$
 \item group generation as in \S \ref{orto} for $n=17,19,21,22,23.$
\end{itemize}

\clearpage
\begin{landscape}
\begin{table}[ht]
\centering \resizebox{1.4\textheight}{!}{
\begin{tabular}{|c||c|c|c|c|c|c|c|c|c|c|c|c|c|c|c|c|c|c|c|c|c|c|c|c|c|c|c|c|c|c|}
  \hline
  \textbf{n/d} & \textbf{1} & \textbf{2} & \textbf{3} & \textbf{4} & \textbf{5} & \textbf{6} & \textbf{7} & \textbf{8} & \textbf{9} & \textbf{10} & \textbf{11} & \textbf{12} & \textbf{13} & \textbf{14} & \textbf{15} & \textbf{16} & \textbf{17} & \textbf{18} & \textbf{19} & \textbf{20} & \textbf{21} & \textbf{22} & \textbf{23} & \textbf{24} & \textbf{25} & \textbf{26} & \textbf{27} & \textbf{28} & \textbf{29} & \textbf{30}
  \\\hline \hline
  \textbf{1} & 1 &  &  &  &  &  &  &  &  &  &  &  &  &  &  &  &  &  &  &  &  &  &  &  &  &  &  &  &  &  \\\hline
  \textbf{2} & 2 & 0 &  &  &  &  &  &  &  &  &  &  &  &  &  &  &  &  &  &  &  &  &  &  &  &  &  &  &  &  \\\hline
  \textbf{3} & 3 & 2 & 1 &  &  &  &  &  &  &  &  &  &  &  &  &  &  &  &  &  &  &  &  &  &  &  &  &  &  &  \\\hline
  \textbf{4} & 4 & 2 &  &0  &  &  &  &  &  &  &  &  &  &  &  &  &  &  &  &  &  &  &  &  &  &  &  &  &  &  \\\hline
  \textbf{5} & 5 & 4 & 1 &  & 1 &  &  &  &  &  &  &  &  &  &  &  &  &  &  &  &  &  &  &  &  &  &  &  &  &  \\\hline
  \textbf{6} & 6 & 4 &  2& 2 &  & 0 &  &  &  &  &  &  &  &  &  &  &  &  &  &  &  &  &  &  &  &  &  &  &  &  \\\hline
  \textbf{7} & 7 & 6 & & 2 &  &  &  1&  &  &  &  &  &  &  &  &  &  &  &  &  &  &  &  &  &  &  &  &  &  &  \\\hline
  \textbf{8} & 8 & 6 &  &  & &  &  &  0&  &  &  &  &  &  &  &  &  &  &  &  &  &  &  &  &  &  &  &  &  &  \\\hline
  \textbf{9} & 9 & 8 & 4 &  & & & &  & 1 &  &  &  &  &  &  &  &  &  &  &  &  &  &  &  &  &  &  &  &  &  \\\hline
  \textbf{10} & 10 & 8 &  &  3& &  & &  &  &0  &  &  &  &  &  &  &  &  &  &  &  &  &  &  &  &  &  &  &  &  \\\hline
  \textbf{11} & 11 & 10 & 5 & 2 & &  & &  &  &  & 1 &  &  &  &  &  &  &  &  &  &  &  &  &  &  &  &  &  &  &  \\\hline
  \textbf{12} & 12 & 10 & 6 & 4 &  &  &  &  &  &  &  &0 &  &  &  &  &  &  &  &  &  &  &  &  &  &  &  &  &  &  \\\hline
  \textbf{13} & 13 & 12 & 6 &  5&  &  &  &  &  &  &  &  & 1 &  &  &  &  &  &  &  &  &  &  &  &  &  &  &  &  &  \\\hline
  \textbf{14} & 14 & 12&  9&  7&  4&  &2 &  &  &  &  &  &  & 0 &  &  &  &  &  &  &  &  &  &  &  &  &  &  &  &  \\\hline
  \textbf{15} & 15 & 14 & 5 & 4 & 4 &  & 2 &  &  &  & &  &  &  & 1 &  &  &  &  &  &  &  &  &  &  &  &  &  &  &  \\\hline
  \textbf{16} & 16 & 14 & 10 &7 &5 &2  & &  &  &  &  &  &  &  &  & 0 &  &  &  &  &  &  &  &  &  &  &  &  &  &  \\\hline
  \textbf{17} & 17 & 16 &  7&  7& 6 &2  &  &  &  &  &  &  &  & & &  & 1 &  &  &  &  &  &  &  &  &  &  &  &  &  \\\hline
  \textbf{18} & 18 & 16 & & 8 & 5 &3  &  &  & &  &  &  &  &  &  &  &  & 0  &  &  &  &  &  &  &  &  &  &  &  &  \\\hline
  \textbf{19} & 19 & 18 &6  &  &  &2 & &  &  & &  &  &  &  &  &  &  &  & 1 &  &  &  &  &  &  &  &  &  &  &  \\\hline
  \textbf{20} & 20 & 18 & 11 &8  &  &5  &  &4 &3  &  &  &  &  &  &  &  &  &  &  & 0 &  &  &  &  &  &  &  &  &  &  \\\hline
  \textbf{21} & 21 & 20 &  &  &4 &  &  &  & 2 &  &  &  & &  &  &  &  &  & &  & 1 &  &  &  &  &  &  &  &  &  \\\hline
  \textbf{22} & 22 & 20 &14  & 12 &7  &  & &4 &  & & 2 &  &  &  &  &  &  &  &  &  &  & 0 &  &  &  &  &  &  &  &  \\\hline
  \textbf{23} & 23 & 22 & 13 & 9 &  & 6 &5  & 3 & &  & & &  &  &  &  & & &  &  & &  & 1 &  &  &  &  &  &  &  \\\hline
  \textbf{24} & 24 & 22 &  16& 14 & 11 & 9 &  8& 7 & 4 & 2 &  &  &  &  & &  &  &  &  &  &  &  &  & 0 &  &  &  &  &  &  \\\hline
   \hline
\end{tabular}}
\caption{
            Lower bounds on LCK$[n,d]$
                    }
\end{table}
\end{landscape}

%%%%%%%%%%%%%%%%%%%%  references  %%%%%%%%%%%%%%%%%%%%%%
\section{Conclusion and open problems}
This paper is dedicated to LCD codes. A linear programming bound on the largest size of an LCD code of given length and distance was derived. It is
a worthwhile project to derive an asymptotic version of that bound. More generally semi-definite programming bounds are worth exploring. A construction based on orthogonal matrices was derived. It would be interesting to see other
classes of combinatorial matrices enter the problem. Improving the table of lower bounds by using codes over rings or symmetric designs is also worth considering.

\end{document}